\documentclass[aps,prb,reprint,floatfix,citeautoscript,longbibliography,superscriptaddress]{revtex4-1}
\usepackage{graphicx,float}
\usepackage{bm,amsmath,amssymb,mathrsfs,dcolumn}
\usepackage{gensymb}
\usepackage[usenames]{color}
\usepackage{subfigure}
\usepackage{ulem}
\usepackage{soul,xcolor}
\usepackage{hyperref}
\hypersetup{colorlinks=true, linkcolor=blue, citecolor=blue, urlcolor=blue}
\setstcolor{red}

\begin{document}

\title{Unconventional magnon excitation by off-resonant microwaves}
\author{H. Y. Yuan}
\affiliation{Institute for Theoretical Physics, Utrecht University, 3584 CC Utrecht, The Netherlands}
\author{Shasha Zheng}
\affiliation{State Key Laboratory for Mesoscopic Physics, School of Physics $\&$ Collaborative Innovation Center of Quantum Matter, Peking University, Beijing 100871, China}
\author{Q. Y. He}
\email[]{qiongyihe@pku.edu.cn}
\affiliation{State Key Laboratory for Mesoscopic Physics, School of Physics $\&$ Collaborative Innovation Center of Quantum Matter, Peking University, Beijing 100871, China}
\author{Jiang Xiao}
\affiliation{Department of Physics and State Key Laboratory of Surface Physics, Fudan University, Shanghai 200433, China}
\author{Rembert A. Duine}
\email[]{r.a.duine@uu.nl}
\affiliation{Institute for Theoretical Physics, Utrecht University, 3584 CC Utrecht, The Netherlands}
\affiliation{Center for Quantum Spintronics, Department of Physics, Norwegian University of Science and Technology, NO-7491 Trondheim, Norway}
\affiliation{Department of Applied Physics, Eindhoven University of Technology, P.O. Box 513, 5600 MB Eindhoven, The Netherlands}
\date{\today}

\begin{abstract}
It is widely recognized that a physical system can only respond to a periodic driving significantly when the driving frequency matches the normal mode frequency of the system, which leads to resonance. Off-resonant phenomena are rarely considered because of the difficulty to realize strong coupling between physical systems and off-resonant waves. Here we examine the response of a magnetic system to squeezed light and surprisingly find that the magnons are maximally excited when the effective driving frequency is several orders of magnitude larger than the resonant frequency. The generated magnons are squeezed which brings the advantage of tunable squeezing through an external magnetic field. Furthermore, we demonstrate that such off-resonant quasi-particle excitation is universal in all the hybrid systems in which the coherent and parametric interaction of bosons exists and that it is purely a quantum effect, which is rooted in the quantum fluctuations of particles in the squeezed vacuum. Our findings may provide an unconventional route to study off-resonant phenomena and may further benefit the use of hybrid matter-light systems in continuous variable quantum information.
\end{abstract}

\maketitle

\section{Introduction}
Resonance is a widely studied phenomenon that a physical system responds to periodic external driving significantly only when the driving frequency matches the normal mode frequency of the system. It was originally found in acoustics as sympathetic resonance where a string vibrates and produces various sounds. By now it has been observed and utilized in many different types of waves, such as mechanical resonance, electromagnetic resonance in abundant classical resonators and waveguides \cite{Jackson}, electron spin resonance \cite{Zavoisky1945}, ferromagnetic and antiferromagnetic resonance \cite{Kittel1948}. Off-resonant phenomena are important to synthesize and manipulate the working of a broad frequency ranges of electromagnetic waves, but they are usually out of concern because of the lack of knowledge or knobs that underpin the concept, until the discovery of nonlinear optical phenomena \cite{DFWalls,shen}. For example, a cavity field can be coupled to a mechanical oscillator through a radiation-pressure like interaction, which allows the direct control of the mechanical mode by tuning the frequency difference of the laser and the cavity mode, even though the mechanical mode is largely off-resonant with the cavity mode \cite{Asp2014}. Recently, this idea was extended to manipulate GHz dynamics of ferromagnets by
optical waves \cite{Cao2020}. Alternatively, if one enhances the coupling of matter and light to the ultra/deep strong coupling regime \cite{Dicke1954,Hepp1973}, then the atomic systems such as a collection of two-level atoms will be strongly excited even at the off-resonant condition. However, this is very challenging to be realized. Here we will take a hybrid magnet-light system as an example to show that the off-resonant physics can naturally manifest under the interplay of squeezed photons and magnets. In principle, our results can be generalized to other hybrid systems with coexistence of coherent and parametric coupling, such as the microwave cavity-magnets with Kerr nonlinearities \cite{Zhang2019} and optomechanical systems \cite{Gan2019,Mar2014}, which couples the cavity field to various high-quality mechanical oscillators.

On the other side, magnons are the elementary excitations in ordered magnets and their phase, amplitude and angular momentum could carry useful information for computing science \cite{Chumak2015}. How to excite, propagate and detect magnons is a key topic in spintronics. A well-known knob to excite magnons is using a microwave field or alternating electric current \cite{Kittel1948,Liu2011,Bracher2017,Demo2006}, where the magnons will be resonantly generated when the driving frequency is close to the natural frequency of the magnons. The resulted magnons are coherent in nature and they have been utilized to drive magnetic textures \cite{PYan2011,Jiang2020}, switch a ferromagnet state \cite{Fukami2016,Yin2018} and can be further converted to electric signals for applications. The off-resonant techniques to excite magnons include femtosecond lasers \cite{Beau1996,Kimel2005,Nishitani2010,Deb2019}, Raman scattering \cite{Fleury1968,Jin2018} and acoustic pulses \cite{Sch2010}.

Besides, cavity magnonics emerges recently to extend the horizon of traditional magnonics, aiming to manipulate the hybrid magnon-photon system near its resonance frequency for quantum information transfer \cite{Soykal2010,Huebl2013, Gor2014,Zhang2014,Wang2018,Cao2015, Vahram2018,Harder2018,Yu2019,Dany2020,Wolz2020,Zhang2019,Yuan2020}.
Here most of the energy level spectrum can be well understood within classical electromagnetism by combining the Maxwell theory and classical magnetization dynamics described by Landau-Lifshitz-Gilbert (LLG) equation \cite{Cao2015,Yu2019}. Whether such a hybrid system can exhibit some quantum properties that have no classical counterpart is, however, not known. This may be of particular importance as one intends to integrate magnonics with quantum information science.

In this article, we study the interaction of squeezed light and a magnonic system and find that magnons are maximally excited far below the effective resonance frequency. The maximum magnon population depends strongly on the squeezing parameter of the light, while it is further tunable through an external magnetic field. The underlying physics is well understood as the interplay of coherent and parametric interaction in such a hybrid system with energy level mismatch. We further ascertain this to be a purely quantum effect, and show how the tunable squeezing of magnons in both quadrature directions can be useful for interpreting the asymmetric steering of magnons and photons.

\begin{figure}
\centering
\includegraphics[width=0.45\textwidth]{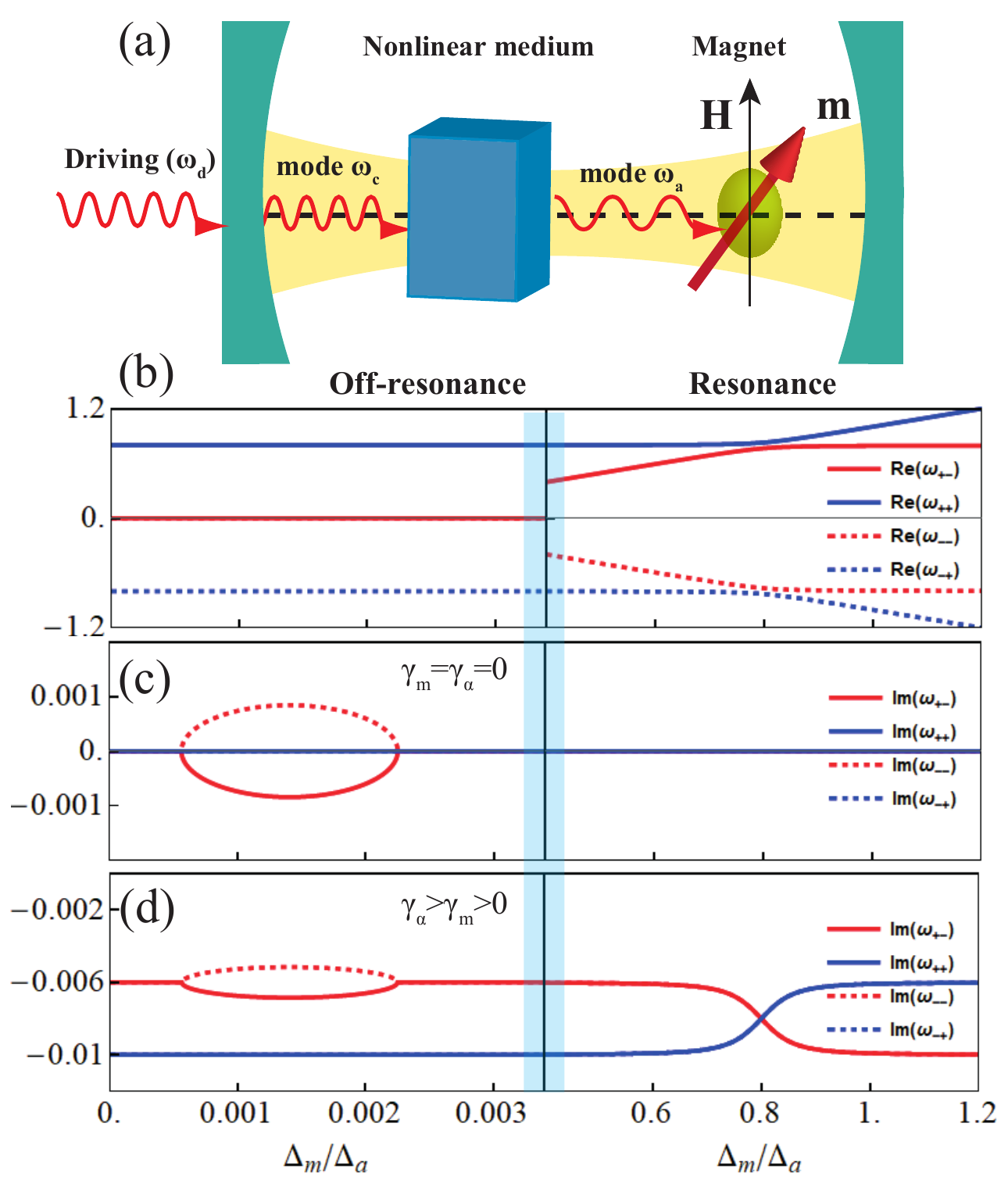}
\caption{(a) Schematic of a magnetic particle interacting with a squeezed light generated by a nonlinear medium. (b) Energy spectrum of the system as a function of magnon frequency. Parameters are $g=0.03\Delta_a$, $\epsilon=0.3\Delta_a$, $\gamma_\alpha=\gamma_m=0$ for (b)(c) and $\gamma_\alpha=0.01,\gamma_m=0.006$ for (d).}
\label{fig1}
\end{figure}

\section{Physical model}
We consider a hybrid magnet-light system shown in Fig. \ref{fig1}(a). A pumping laser with frequency $\omega_d$ interacts with a nonlinear medium and creates pair of squeezed photons with frequency $\omega_a$ through a parametric down conversion process. These squeezed photons interact with the magnet via Zeeman interaction. The minimal Hamiltonian describing such a process can be written as,
\begin{equation}
\begin{aligned}
\mathcal{H}&=\omega_c c^\dagger c + \omega_a a^\dagger a +  \omega_m m^\dagger m + \epsilon' (ca^\dagger a^\dagger + c^\dagger aa) \\
&  + g(a^\dagger m+a m^\dagger) + \zeta(c^\dagger e^{-i\omega_d t}+c e^{i\omega_d t}),
\end{aligned}
\end{equation}
where $c,a,m$ are the annihilation operators for laser photons, the subharmonic photons and the magnons, respectively. Here, $\epsilon'$ is the interaction strength between the laser and subharmonic photons, which is proportional to the (effective) second-order susceptibility of the nonlinear medium, and $g$ is the coupling strength between subharmonic photons and magnons. Here the parametric coupling term between magnon and photon $a^\dagger m^\dagger +a m$ is dropped under the rotating wave approximation (RWA) \cite{Asp2014} because $g\ll \omega_a,\omega_m$ and it only contributes a fast oscillation effect with the frequency $\omega_a+\omega_m$. In a rotating frame, we can adiabatically eliminate mode $c$ if it possesses a large detuning or dissipation and recast the effective Hamiltonian as,
\begin{equation}
	\mathcal{H}=\Delta_a a^\dagger a + \Delta_m m^\dagger m + \epsilon (a^\dagger a^\dagger + aa) + g(a^\dagger m+am^\dagger),
	\label{Hamk}
\end{equation}
where $\Delta_a=\omega_a-\omega_d/2,\Delta_m=\omega_m-\omega_d/2$, and where $c$ is replaced by its average value under strong driving, i.e. $\epsilon=\epsilon'\langle c \rangle$. Note that both the photon and magnon modes ($a$ and $m$) are quantum operators that satisfy commutation relations, instead of classical $c$-numbers.

The Hamiltonian (\ref{Hamk}) can be completely diagonalized by standard techniques \cite{Xiao2009}. To see, however, the influence of squeezed photons on the eigenspectrum as well as the magnon population, it is useful to first diagonalize the photonic part of (\ref{Hamk}) by introducing a squeezed operator $a=\cosh \theta \alpha  + \sinh \theta \alpha^\dagger$ after which we obtain,
\begin{equation}
\mathcal{H}=\Delta_\alpha \alpha^\dagger \alpha + \Delta_m m^\dagger m + g_c(\alpha^\dagger m+\alpha m^\dagger) + g_s (\alpha^\dagger m^\dagger + \alpha m),
\label{Hamk2}
\end{equation}
where $\Delta_\alpha=\sqrt{\Delta_a^2-4\epsilon^2}$ is the modified photon frequency, $g_c=g \cosh\theta$, $g_s=g\sinh \theta$, and $\theta=\mathrm{arctanh}(-2\epsilon/\Delta_a)/2$ is the squeezing parameter of the photons. The Hamiltonian (\ref{Hamk2}) shows that the normal mode ($\alpha$) can not only convert photons to magnons through a coherent coupling $g_c$, but also excite magnons and photons simultaneously through a parametric coupling $g_s$. Now the parametric term cannot be neglected, because, as we shall see, the most interesting physics occurs when $\Delta_m <g_s\ll \Delta_\alpha$,
where the RWA cannot be made (See Appendix A for more justification of this Hamiltonian).

The Heisenberg equation of magnon and photon mode, governed by the Hamiltonian (\ref{Hamk2}) can be written as $d\mathbf{R}/dt=\mathbf{M} \cdot \mathbf{R}$ with $\mathbf{R}=(x_m,p_m,x_\alpha,p_\alpha)^T$, $x_\nu=(\nu+\nu^\dagger)/\sqrt{2},p_\nu=(\nu-\nu^\dagger)/\sqrt{2}i$, with $\nu=\alpha, m$ and
\begin{equation}
\mathbf{M}=\left ( \begin{array}{cccc}
            -\gamma_m & \Delta_m &0&g_c-g_s \\
            -\Delta_m & -\gamma_m &-(g_c+g_s)&0 \\
            0 & g_c-g_s &-\gamma_\alpha&\Delta_\alpha \\
            -(g_c+g_s) & 0 &-\Delta_\alpha&-\gamma_\alpha \\
          \end{array}
\right ),
\end{equation}
where the dissipation has been introduced as $\omega_\nu \rightarrow \omega_\nu-i\gamma_\nu$. Using the ansatz $\mathbf{R}(t)=e^{-i\omega t} \mathbf{R}$, the eigenspectrum of the system can be readily determined by solving the secular equation $\det(\omega -i \mathbf{M})=0$. When $\gamma_m=\gamma_\alpha=\gamma$, the spectrum is analytically calculated as,
\begin{equation}
\begin{aligned}
&\omega_{\pm\pm }= \pm\left \{ \frac{1}{2}(\Delta_m^2+\Delta_\alpha^2) +  g^2 \pm \frac{1}{2}\left [ 16g^2 \Delta_m \Delta_\alpha \sinh^2 \theta \right . \right .\\
&\left . \left. ~~~~~ +(\Delta_m+\Delta_\alpha)^2((\Delta_m-\Delta_\alpha)^2 + 4g^2) \right ] ^{1/2} \right \}^{1/2}-i\gamma.
\end{aligned}
\label{eigen}
\end{equation}
When $\theta=0,\gamma=0$, one recovers the well-studied energy level repulsion spectrum in the coherent limit (real $g$) and the level attraction spectrum in the dissipative limit (imaginary $g$). Figure \ref{fig1}(b) shows the energy spectrum of the system. A level repulsion appears when the magnons become resonant with the photons, when $\Delta_m=\Delta_\alpha=0.8\Delta_a$. Besides, an anomalous valley in the magnonic branch is identified at a magnon frequency far from resonance ($\Delta_m \ll \Delta_\alpha$), as shown in Fig. \ref{fig1}(c) (red line). Here, eigenfrequency becomes imaginary with a maximum at a particular value of detuning, even if the hybrid system is free from dissipation ($\gamma_m=\gamma_\alpha=0$). Mathematically, we describe this anomaly by expanding the magnonic branch in Eq. (\ref{eigen}) around a small value of $\Delta_m$ and find that $\omega_{\pm-}$ becomes complex when
\begin{equation}
\frac{g^2}{\Delta_\alpha}I_- (\theta)<\Delta_m< \frac{g^2}{\Delta_\alpha}I_+ (\theta),
\label{regime}
\end{equation}
where $I_\pm (\theta) =2\cosh 2\theta \pm \sqrt{2\cosh 4 \theta}$. To guarantee the stability of the system, the imaginary parts of all the eigenvalues should be negative, which further requires,
\begin{equation}
\gamma > \sqrt{(g^2 I_- (\theta)-\Delta_m \Delta_\alpha )(\Delta_m \Delta_\alpha -g^2 I_+ (\theta))/2\Delta_\alpha^2}.
\label{rhc}
\end{equation}
When $\gamma_m \neq \gamma_\alpha$, the features of the spectrum at low magnon frequency are similar to the case of $\gamma_m = \gamma_\alpha$, except a typical level crossing appears near the resonance, as shown in Fig. \ref{fig1}(d).

\section{Unconventional magnon excitation}
Generally, an enhanced dissipation of magnons may lead to a reduction of magnon number. However, it is the other way around here. To illustrate this point, we start from the Heisenberg equation for the quadratic operators of the hybrid system,
\begin{equation}
d\mathbf{u}/dt=\mathbf{D}\cdot \mathbf{u} +\mathbf{v},
\label{heisen}
\end{equation}
where $\mathbf{u}=(\langle \alpha^\dagger \alpha \rangle,\langle m^\dagger m \rangle,\langle \alpha m \rangle,\langle mm \rangle,\langle \alpha \alpha \rangle,\langle \alpha^\dagger m \rangle)$, $\mathbf{D}$ is a drift matrix, $\mathbf{v}=(0,0,-ig_s,0,0,0)$. The steady-state solution of Eq. (\ref{heisen}) exists only when all the real parts of eigenvalues of $\mathbf{D}$ are negative. This sets a stability condition for the system, which is consistent with Eq. (\ref{rhc}). Unless otherwise noted, we take parameters in the stable regime throughout this article.  By setting $d\mathbf{u}/dt=0$, we can analytically obtain the steady-state occupations of magnons $\langle m^\dagger m \rangle$ and photons $\langle \alpha^\dagger \alpha \rangle$ (See Appendix B for the analytical expression).

\begin{figure}
	\centering
	\includegraphics[width=0.48\textwidth]{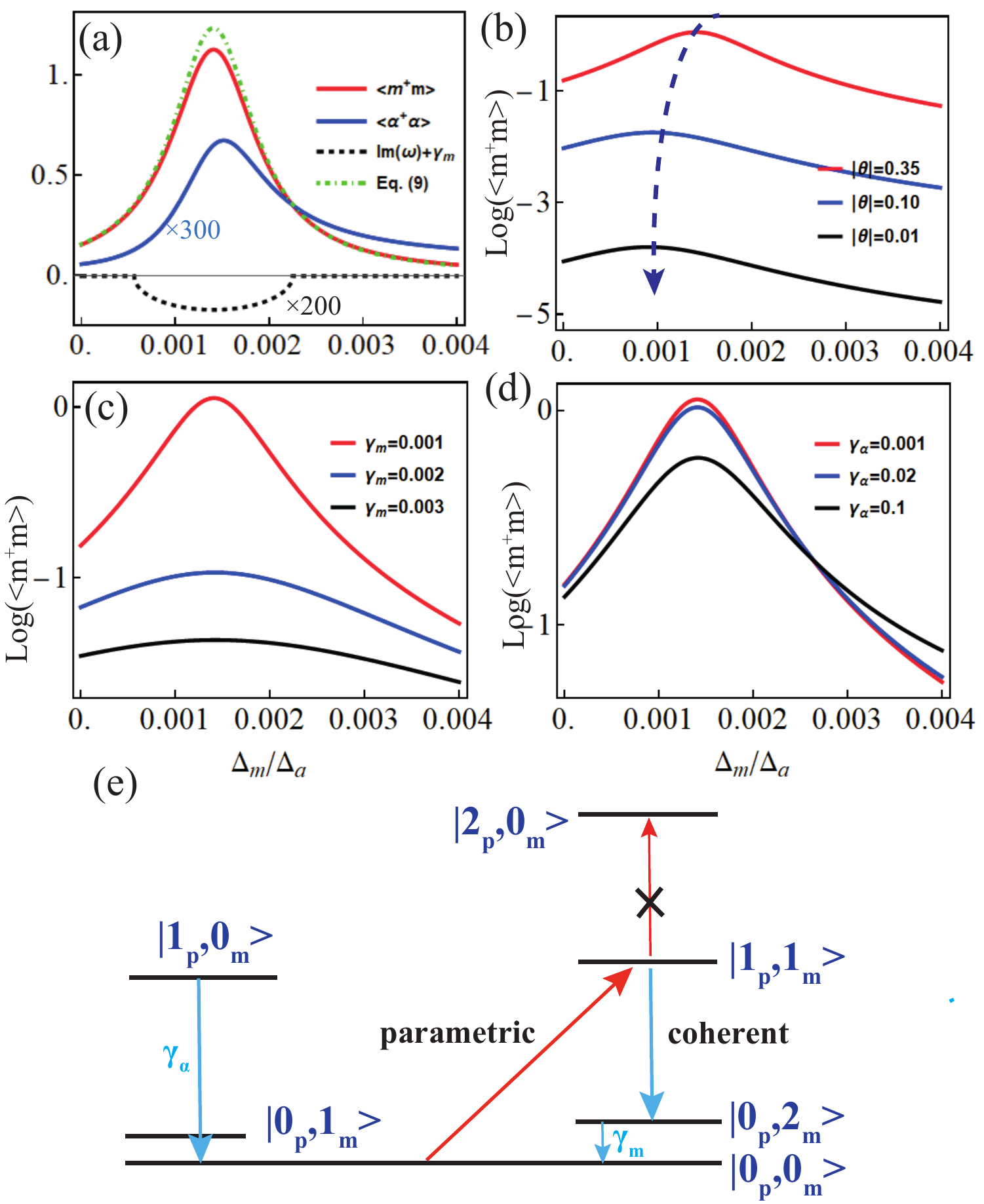}
	\caption{(a) Magnon (red line) and photon (blue line) populations as a function of magnon frequency near the anomalous points. (b) Magnon population as a function of the magnon frequency for squeeze parameter $|\theta|=0.35$ (red line), 0.10 (blue line), and 0.01 (black line), respectively. $\gamma_\alpha/\Delta_a=0.01,\gamma_m/\Delta_a=0.001,g/\Delta_a=0.03$. (c) Magnon population as a function of magnon frequency at $\gamma_m/\Delta_a=0.001$ (red line), 0.002 (blue line) and 0.003 (black line), respectively. (d)  Magnon population as a function of magnon frequency at $\gamma_\alpha/\Delta_a=0.001$ (red line), 0.02 (blue line) and 0.1 (black line), respectively. (e) Schematic of the physical picture \cite{notepic}. }
	\label{fig2}
\end{figure}

In the limiting case $\gamma_\alpha=0$, the magnon density at the off-resonant condition ($\Delta_\alpha \gg g \gg \Delta_m$) can be approximated as (See Appendix B for the derivation process),
\begin{equation}
\langle m^\dagger m \rangle \approx \frac{2g_c^2g_s^2}{g^4+\gamma_m^2\Delta_\alpha^2+\Delta_\alpha^2(\Delta_m^2-2g^2\Delta_m \mathrm{cosh}(2\theta)/\Delta_\alpha)}.
\label{approx}
\end{equation}
One immediately sees that the magnons are maximally excited at the point $\Delta_m=g^2\mathrm{cosh}2\theta /\Delta_\alpha$, which falls into the regime specified by Eq. (\ref{regime}). Eq. (\ref{approx}) also shows that the magnon density may diverge when the dissipation in the denominator is not large enough to cancel the effect by parametric pumping above the critical magnon frequency.
To verify these predictions, we plot the accurate magnon and photon number as a function of magnon frequency in Fig. \ref{fig2}. (i) A maximum magnon and photon population occurs at the far off-resonance with $\Delta_m/\Delta_a=1.4 \times 10^{-3}$, coincident with the maximum enhancement damping of the system as shown by the dashed line in Fig. \ref{fig2}(a). (ii) The maximum magnon population decreases as the squeezing parameter of the light decreases as shown in Fig. \ref{fig2}(b), and finally approaches zero for a non-squeezed light. (iii) There exists a threshold of magnetic dissipation $8 \times 10^{-4}$, below which the system becomes unstable, indicating that the magnetic system reaches the non-linear regime. Above the threshold, the maximum magnon occupation decreases as the damping increases, as shown in Fig. \ref{fig2}(c). All these features are consistent with the predictions of Eq. (\ref{approx}) quantitatively. Moreover, (iv) The magnon population is three orders of magnitude larger than the photon populations, its magnitude is insensitive to the photon dissipation, up to $\gamma_\alpha=0.02$, as shown in Fig. \ref{fig2}(d).

To understand the essential physics intuitively, we clarify three channels that contribute to the magnon population. (i) The dissipation of magnons subjects to intrinsic damping $\gamma_m$; (ii) The parametric excitation of magnons and photons simultaneously due to the process $\alpha^\dagger m^\dagger$. This is the major source to produce the finite occupation of magnons and photons; (iii) The coherent pumping of magnons from the photon system through the beam-splitter-type interaction $g_c$. To see how the system builds a steady state, we first sketch the energy level diagram of the hybrid system in Fig. \ref{fig2}(e). Note that the off-resonant condition $\Delta_m/\Delta_a \ll 1$ implies that the energy levels spacing of photons is much larger than that of magnons. Starting from a vacuum state $|0_p, 0_m \rangle$, a pair of magnon and photon is first excited to the level $|1_p, 1_m \rangle$ parametrically ($g_s$). Now it can either evolve to the state $|2_p, 0_m \rangle$ or $|0_p, 2_m \rangle$ through the coherent particle transfer ($g_c$). Since $\Delta_m/\Delta_a \ll 1$, the $\alpha^\dagger m$ process will be energetically consuming and thus the $\alpha m^\dagger$ process will dominate to generate the state $|0_p, 2_m \rangle$. Therefore, the magnons will be accumulated. Throughout the process, all the Fock states will keep dissipating to the ground state ($\gamma_m$).
In total, the pumping, transfer and dissipation process compete and finally reach a fine balance, building a steady-state magnon distribution (See Appendix C for a generalization of this intuitive picture). From the view of angular momentum conservation, the pumping laser is the angular momentum source to conserve the total angular momentum of the system. As a comparison, when the light is not squeezed ($\theta=0$), parametric pumping ($g_s$) is absent, the total number of magnons and photons will decrease to zero gradually due to the intrinsic dissipation.



\section{Squeezed magnons}
We have shown that magnons can be maximally excited by a squeezed light at the off-resonant condition. It would be meaningful to see whether the quantum information of photons such as their squeezing can be transferred to magnons at off-resonant conditions, since the magnon squeezing is a useful resource in quantum physics \cite{Kamra2016, Sharma2020,Brataas2020}. To answer this question, we evaluate the uncertainty of the quadrature of magnons and photons as,
$\langle \Delta x_\nu^2 \rangle = n_\nu + \Re(\langle \nu\nu \rangle)+\frac{1}{2}$,
$\langle \Delta p_\nu^2 \rangle = n_\nu - \Re(\langle \nu\nu \rangle)+\frac{1}{2}$, where $\nu=\alpha,m$, $\Re(\langle \nu\nu \rangle)$ is the real part of $\langle \nu\nu \rangle$, which has already been derived in solving Eq. (\ref{heisen}).
\begin{figure}
	\centering
	\includegraphics[width=0.48\textwidth]{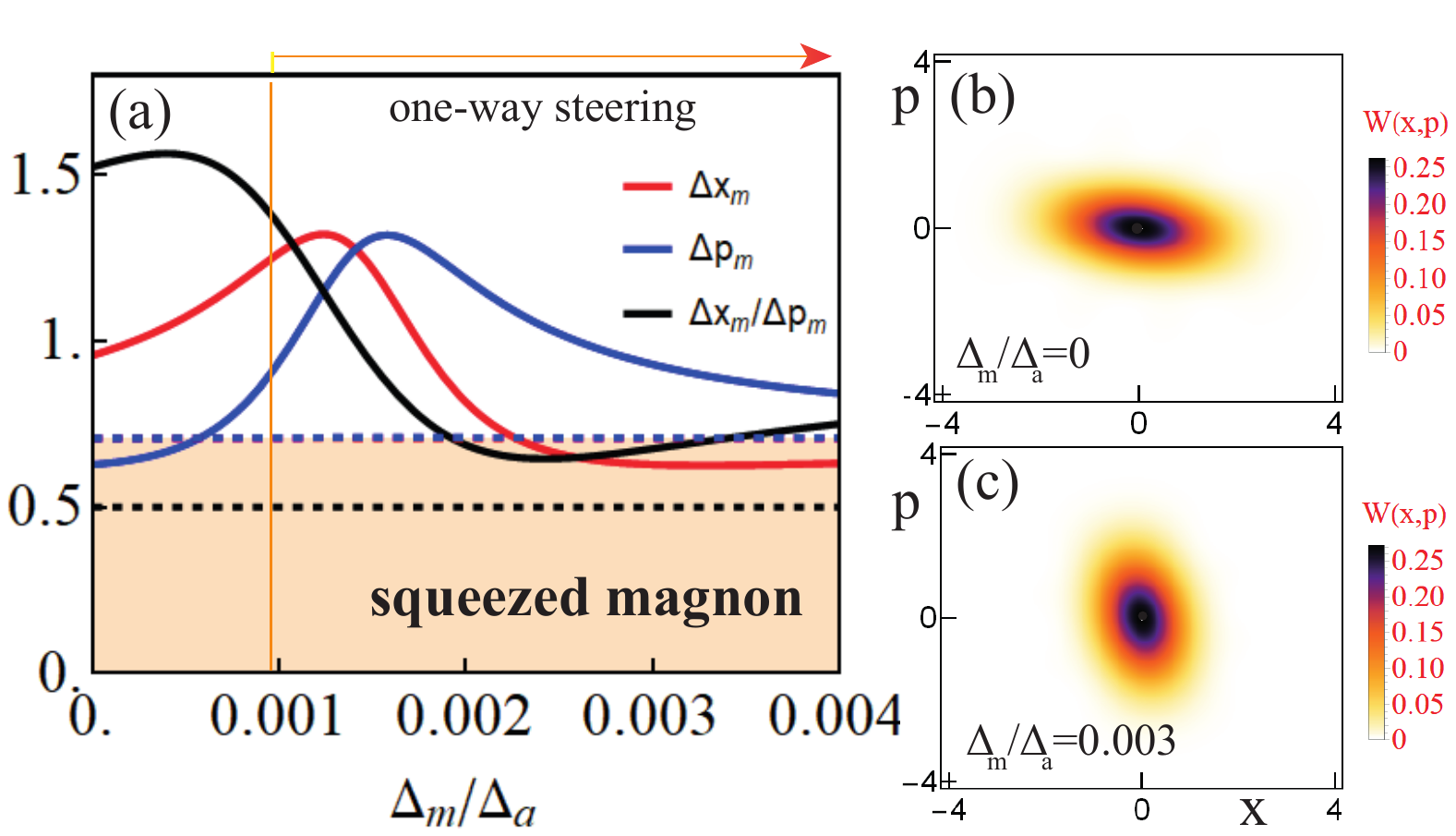}
	\caption{(a) Uncertainty of the quadratures of magnons as a function of magnon frequency. The dashed line at $1/\sqrt{2}$ is the standard quantum limit, below which a state becomes squeezed. The dashed line at 0.5 is the squeeze parameter of the input light ($\alpha$ mode). (b)(c) shows the Wigner distribution of the steady states at $\Delta_m/\Delta_a=0$ and 0.003. All the parameters are the same as Fig. \ref{fig2}(a).}
	\label{fig3}
\end{figure}
Figure \ref{fig3}(a) shows the uncertainty $\Delta x_\nu \equiv\sqrt{\langle \Delta x_\nu^2 \rangle},\Delta p_\nu \equiv \sqrt{\langle \Delta p_\nu^2 \rangle}$ as a function of magnon frequency. Clearly, the uncertainty of magnon quadrature $\Delta p_m$ ($\Delta x_m$) is squeezed below the quantum limit $1/\sqrt{2}$ when $\Delta_m/\Delta_a$ has a red (blue) shift from the off-resonant peak, as illustrated in Fig. \ref{fig3}(b) and \ref{fig3}(c) \cite{note1}. This suggests that the squeezing of the light is transferred to the magnons. As the magnon frequency approaches the resonant condition at $\Delta_m=\Delta_\alpha$, no significant squeezing of magnons is observed.

The tunable squeezing property of magnons through the external static field may imply the one-way quantum steering of photons by detecting magnons. Here we quantify the steering properties between magnons and photons (See Appendix D for the numerical method to quantity the steering), and find that magnons can steer the photon state when the magnons are tuned to be squeezed in the $x$ direction, as indicated in Fig. \ref{fig3}(a). The reversed steering process from photon to magnon is suppressed, due to the particle number asymmetry of the excitations. When the magnons are squeezed in the $p$ direction, this steering channel between magnons and photons are closed.


%

\section{Discussions and Conclusions}
To realize our predictions, we propose to couple a low damping magnetic insulator such as yttrium iron garnet to a cavity with nonlinear medium that supports microwave parametric down conversion \cite{Moon2005,Cao2011,Wang2015}. Here the resonant frequency of ferromagnet ($\omega_m$) is several GHz with inverse quality factor ranging from $10^{-5}\sim 10^{-2}$, and the coupling strength between magnon and microwave ($g$) ranges from tens of MHz up to gigahertz \cite{Huebl2013,Gor2014,Bai2015,Wang2018,Harder2018,Bohi2019,YPWang2019,Wolz2020}. Hence the parameter regime of our theoretical proposal is accessible in the experiments (See Appendix E for a list of the experimental parameters). We expect a sudden broadening of the transmission linewidth and a maximal excitation of magnons when the magnon frequency is tuned to the anomalous regime. This is different from the zero damping phenomena reported in non-Hermitian magnet-light systems \cite{YPWang2019}. The generated squeezed state can be further measured by another probe field \cite{Sharma2020}.


We notice that a significant number of two-mode systems in classical physics can be quantized into a form resembling Hamiltonian (\ref{Hamk2}). The first class is the two classical harmonic oscillators coupled through a spring, i.e. $\mathcal{H}_{cl}=\frac{1}{2}\sum_{i=1}^2(m_i\dot{x}_i^2 + m_i\omega_i^2 x_i^2)+k(x_1-x_2)^2$,
where $m_i$ is the mass, $\omega_i$ is the intrinsic frequency of the oscillators, and $k$ is the spring stiffness. The spectrum of this system, however, does not show the anomaly presented here, because the effective coupling $k$ modifies the natural frequency of the modes to invalidate the condition (\ref{regime}). An artificial model that simplifies the coupling as $-2k x_1 x_2$ can avoid the modification of the natural frequency, but it will suffer from a divergence problem for the distinguished description of dissipation in Newtonian mechanics and quantum physics (See Appendix F for the mathematical details). The second class is that of two ferromagnetically coupled spins, i.e. $\mathcal{H}_{cl}=-B_1 S_{1,z} - B_2 S_{2,z} - J S_{1x} S_{2x}$, where $B_1$ and $B_2$ are the applied fields and $J$ is the effective coupling. The ground state of this system will tilt in the anomalous regime ($B_1B_2<J^2$) and the finite excitation disappears around the tilted ground state caused by Gilbert damping (See Appendix F for the mathematical details).

In summary, we have studied the interaction of a nanomagnet with squeezed light and identified a surprising maximum occupation of magnons when the magnon frequency is far below the resonance. The essential physics is understood by the interplay among magnon pumping effects by the squeezed photons, coherent conversion of photons to magnons and the intrinsic magnon loss.
In the anomalous regime, the magnon pumping effect is amplified and leads to an explosion of magnons and the reciprocal process of magnon loss manifests as an enhanced damping of magnons. We further show that such off-resonant excitation of particles is a purely quantum phenomenon, where a classical system will either evolve to a ground state with zero particle fluctuation or become unstable. Our findings provide a generic platform to examine the difference between classical physics theory and quantum mechanics and further benefit the applications of hybrid magnet-light system for quantum information. The intuitive physics picture can be further generalized to study the off-resonant behaviors in a wide class of quantum mechanical systems with the coexistence of coherent and parametric interactions.

\section{Acknowledgments}
This project has received funding from the European Research Council (ERC) under the European Unions Horizon 2020 research and innovation programme (grant agreement No. 725509). RD is member of the D-ITP consortium, a program of the Netherlands Organisation for Scientific Research (NWO) that is funded by the Dutch Ministry of Education, Culture and Science (OCW). Q.Y. H acknowledges the support from NSFC (Grants No. 61675007 and No. 11975026) and the National Key R\&D Program of China (Grants No. 2018YFB1107205 and No. 2016YFA0301302).

\appendix

\section{Why not directly couple magnon with photon?}

The Hamiltonian that describes the direct coupling between a linearly-polarized light and the magnet can be written as,
\begin{equation}
 H=\omega_a a^\dagger a + \omega_m m^\dagger m +g (a+a^\dagger)(m+m^\dagger).
 \label{amdirect}
\end{equation}
Intuitively, one may tune the magnon frequency to a lower value and let the system enter into the anomalous regime $g\sim \sqrt{\omega_a \omega_m}$ discussed in the main text, then the off-resonant excitation will appear. However, this may not be true because Hamiltonian (\ref{amdirect}) by treating the coupling coefficient $g$ as a constant only works near the resonance and with $g \ll \omega_m,\omega_a$ for the following reasonings:(i) a soft magnetic sphere like yttrium iron garnet (YIG) may break into multi-domain in such a low value of magnon frequency that immediately invalids the applicability of the above Hamiltonian. (ii)If we increase the volume of magnet to increase the coupling g, the magnet itself will become a cavity and influence the cavity field significantly. Again, this will invalidate the macrospin approach.

In principle, we need to combine the Maxwell equation and LLG equation to explain the physics in the regime $g\sim \sqrt{\omega_a \omega_m}$, the details of which will be published elsewhere. Here we only give a short introduction on how the spectrum may look like when $\omega_m$ is very small by treating the resonator as a $RLC$ circuit \cite{Bohi2019}. It is still an open question on whether we can find some exotic magnet and cavity system to realize the condition $\omega_m <g \ll \omega_a$, which will be discussed elsewhere. The dynamics of the coupled $RLC$-circuit and magnetic system is described by the circuit equation,
\begin{equation}
L\frac{dj}{dt}+ Rj+\frac{1}{C} \int j dt = V_m,
\label{rlcsm}
\end{equation}
where $L$, $R$ and $C$ are respectively the resistance, inductance and capacitance of the system. The cavity frequency is $\omega_a=1/\sqrt{LC}$ and R describes the strength of dissipation. $V_m$ is the driving force coming from the the precessing magnetization. From the Faraday law, we have $V_{m,x} = i \omega K_cLs_y e^{i \omega t},V_{m,y} = -i \omega K_cLs_x e^{i \omega t}$. From Amp\`{e}re law, we have $h_x = K_m j_y, h_y = -K_m j_x$. Substituting these relations into Eq. (\ref{rlcsm}), we obtain
\begin{equation}
(\omega^2 - 2i\omega \omega_a \gamma_a - \omega_a^2)h^- + \omega^2K_cK_m s^- = 0,
\label{rlc1}
\end{equation}
where $s^-=s_x-is_y,h^-=h_x-ih_y$. By linearizing LLG equation Eq. (\ref{llg}) around the ground state $\mathbf{S}=e_z + (s_x e_x+s_y e_y)e^{i \omega t}$, we have
\begin{equation}
(\omega + i\alpha \omega - \omega_H)s^- + \omega_M h^- = 0.
\label{rlc2}
\end{equation}
where $\omega_M$ corresponds to saturation magnetization. Combining Eqs. (\ref{rlc1}) and (\ref{rlc2}) to have non-zero solutions, it is required that,
\begin{equation}
\left |
\begin{array}{cc}
  \omega + i\alpha \omega - \omega_H & \omega_M \\
  \omega^2K_cK_m & \omega^2 - 2i\omega \omega_a \gamma_a - \omega_a^2
\end{array}
\right | =0.
\label{secular}
\end{equation}

\begin{figure}
\centering
\includegraphics[width=0.48\textwidth]{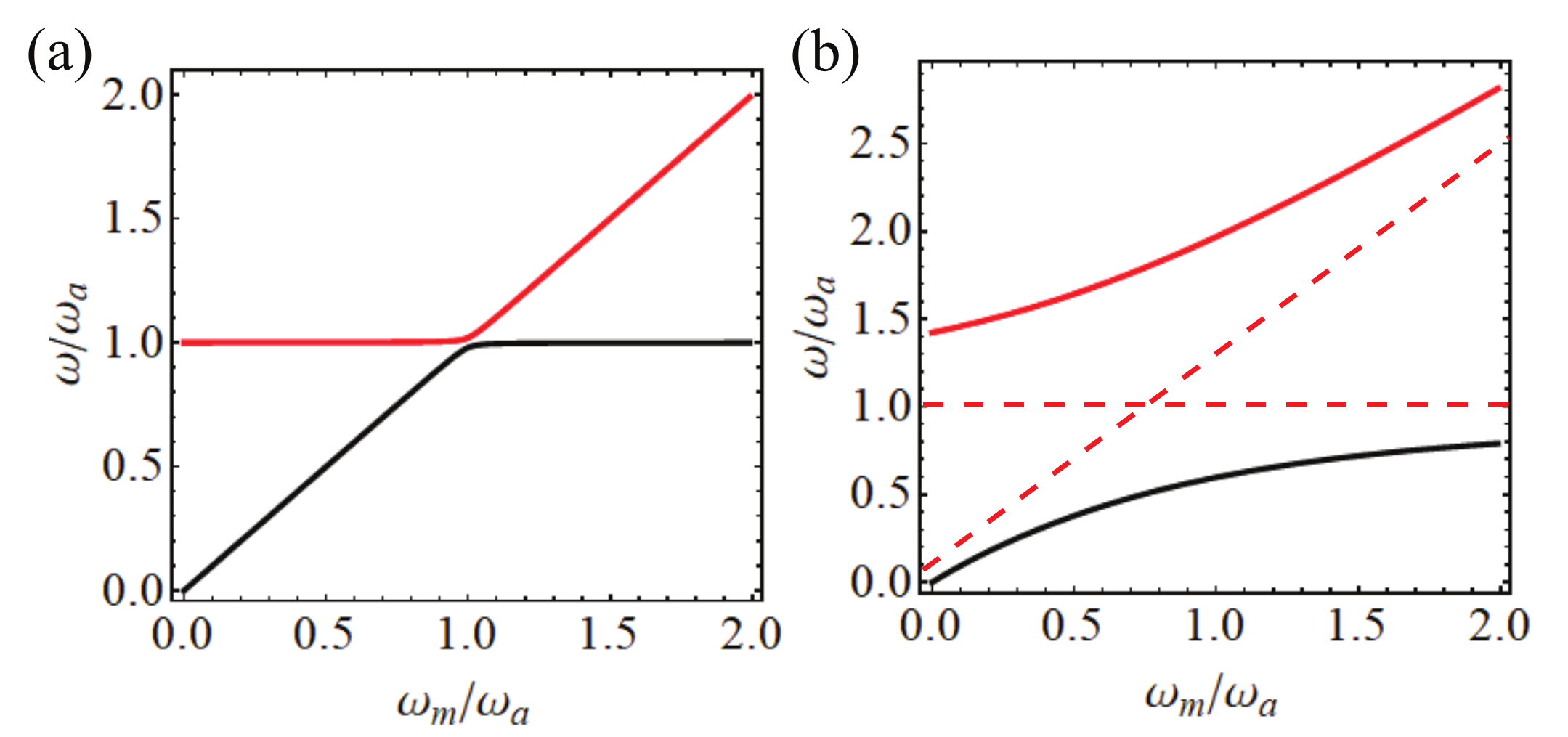}
\caption{Energy spectrum in the $RLC$ circuit when magnon and photon reaches strong coupling (a) and ultrastrong coupling regime (b). Parameters are $\omega_M=\omega_a,\gamma_c=0.01\omega_a$. $K_c=K_m=0.6\sqrt{2}$ for (a) and $K_c=K_m=0.6\sqrt{2}$ for (b).}
\label{rlc}
\end{figure}

In the absence of dissipation($\alpha=0,\gamma_a=0$), the eigenvalues can be solved directly near resonance as,
\begin{equation}
\omega = \frac{\omega_H + \omega_a \pm \sqrt{(\omega_H-\omega_a)^2 +4g_{\mathrm{eff}}^2}}{2(1-g_{\mathrm{eff}}^2/(\omega_H\omega_a))},
\end{equation}
where $g_{\mathrm{eff}}=\sqrt{\omega_M\omega_HK_cK_m/2}$. One immediately see that the effective coupling depends on the
magnon frequency ($\omega_H$), which will guarantee that the eigenvalues are always positive regardless of the coupling strength at resonance. To show how this result is robust at off-resonant condition, we numerically solve Eq. (\ref{secular}) and plot the energy spectrum in Fig. \ref{rlc}. We can see that the spectrum is the normal energy level repulsion, regardless of the strength of coupling.

\section{Magnon and photon occupation in the steady state}

To solve for the steady magnon and photon occupation, we recall the dynamic equations in the Lindblad formalism,
\begin{equation}
\frac{d \langle \mathcal{O} \rangle}{dt}=-i\langle [\mathcal{O},\mathcal{H}] \rangle + \langle \mathcal{D}(\mathcal{O}) \rangle
\label{lindblad}
\end{equation}
where $\mathcal{H}$ is the Hamiltonian of the system, $\langle \mathcal{O} \rangle$ is the ensemble average of the observable $\mathcal{O}$, and
\begin{equation}
\mathcal{D}(\mathcal{O})= \sum_{\nu=\alpha,m}\gamma_\nu \left( \nu^\dagger [\mathcal{O},\nu] + [\nu^\dagger, \mathcal{O}]\nu\right ).
\end{equation}
The steady state $d \langle \mathcal{O} \rangle / dt=0$ implies that
\begin{equation}
-i\langle [\mathcal{O},\mathcal{H}] \rangle + \langle \mathcal{D}(\mathcal{O}) \rangle=0.
\end{equation}
Note that the evolution of particle density $\langle \alpha^\dagger \alpha \rangle,\langle m^\dagger m \rangle$ is always dependent on the evolution of other quadrature operators $\langle \alpha \alpha \rangle,\langle \alpha^\dagger m \rangle,\langle \alpha m \rangle,\langle mm \rangle$. We must solve the dynamics of them in a complete set simultaneously as,
\begin{widetext}
\begin{subequations}
\begin{align}
&g_c Im(\langle \alpha^\dagger m \rangle)-2 g_s Im(\langle \alpha m \rangle)-2 n_\alpha \gamma _{\alpha }=0,\\
&-2 g_c Im(\langle \alpha^\dagger m \rangle)-2 g_s Im(\langle \alpha m \rangle)-2 n_m \gamma _m=0,\\
&g_c (Im(\langle \alpha \alpha \rangle)+Im(\langle mm \rangle)+Im(\alpha m) \left(\Delta _{\alpha }+\Delta _m\right)-Re(\langle \alpha m \rangle) \left(\gamma _{\alpha }+\gamma _m\right)=0,\\
&-\left(g_c (Re(\langle \alpha \alpha \rangle)+Re(\langle m m \rangle)+g_s (n_\alpha+n_m+1)+Re(\langle \alpha m \rangle) \left(\Delta _{\alpha }+\Delta _m\right)\right)-Im(\alpha m) \left(\gamma _{\alpha }+\gamma _m\right)=0,\\
&2 g_c Im(\alpha m)+2 g_s Im(\langle \alpha^\dagger m \rangle)+2 Im(\langle m m \rangle) \Delta _m-2 Re(\langle m m \rangle) \gamma _m=0,\\
&-\left(2 g_c Re(\langle \alpha m \rangle)+2 g_s Re(\langle \alpha^\dagger m \rangle)+2 Re(\langle m m \rangle) \Delta _m\right)-2 Im(\langle m m \rangle) \gamma _m=0,\\
&2 g_c Im(\alpha m)-2 g_s Im(\langle \alpha^\dagger m \rangle)+2 Im(\langle \alpha \alpha \rangle) \Delta _{\alpha }-2 Re(\langle \alpha \alpha \rangle) \gamma _{\alpha }=0,\\
&-\left(2 g_c Re(\langle \alpha m \rangle)+2 g_s Re(\langle \alpha^\dagger m \rangle)+2 Re(\langle \alpha \alpha \rangle) \Delta _{\alpha }\right)-2 Im(\langle \alpha \alpha \rangle) \gamma _{\alpha }=0,\\
&\left(g_s (-Im(\langle \alpha \alpha \rangle)-Im(\langle mm \rangle)+Im(\langle \alpha^\dagger m \rangle) \left(\Delta _m-\Delta _{\alpha }\right)\right)-Re(\langle \alpha^\dagger m \rangle) \left(\gamma _{\alpha }+\gamma _m\right)=0,\\
&-\left(g_c (n_\alpha-n_m)+g_s (Re(\langle \alpha \alpha \rangle)-Re(\langle m m \rangle)+Re(\langle \alpha^\dagger m \rangle) \left(\Delta _m-\Delta _{\alpha }\right)\right)-Im(\langle \alpha^\dagger m \rangle) \left(\gamma _{\alpha }+\gamma _m\right)=0,
\end{align}
\label{leq10}
\end{subequations}
\end{widetext}
where $Re(x)$ and $Im(x)$ represent the real and imaginary parts of the number $x$, respectively.
By solving this set of linear equations, we obtain,
\begin{subequations}
\begin{align}
\langle m^\dagger m \rangle &=\frac{g^2 \sinh^2\theta (A+B+C+D)}{Z_1+Z_2+Z_3+Z_4}, \\
\langle \alpha^\dagger \alpha \rangle &=\frac{g^2 \sinh^2\theta(A'+B'+C'+D')}{Z_1+Z_2+Z_3+Z_4},
\end{align}
\end{subequations}

where
\begin{widetext}
\begin{equation}
\begin{aligned}
A&=\gamma_\alpha  \left(\gamma_\alpha ^2+\Delta_\alpha ^2\right) (\gamma_\alpha +\gamma_m) \left(\gamma_m^2+\Delta_m^2\right) \left((\gamma_\alpha +\gamma_m)^2+(\Delta_m-\Delta_\alpha )^2\right), \\
B&=g^4 (\gamma_\alpha +\gamma_m) \left(-\gamma_\alpha  \Delta_\alpha  \Delta_m+\Delta_\alpha ^2 (\gamma_\alpha +\gamma_m)+\gamma_\alpha  \left((\gamma_\alpha +\gamma_m)^2+\Delta_m^2\right)\right),\\
C&=g^2 \left(\gamma_\alpha ^2 \Delta_\alpha ^3 \Delta_m+\gamma_\alpha  \Delta_\alpha  \Delta_m \left(\gamma_\alpha ^3+\gamma_m^3+\gamma_m \Delta_m^2\right)+\gamma_m \Delta_\alpha ^2 \left(3 \gamma_\alpha ^3+\gamma_m^3+4 \gamma_\alpha  \gamma_m^2+6 \gamma_\alpha ^2 \gamma_m+\gamma_m \Delta_m^2\right) \right .\\
&\left .+2 \gamma_\alpha ^2 \gamma_m (\gamma_\alpha +\gamma_m) \left((\gamma_\alpha +\gamma_m)^2+\Delta_m^2\right)+\gamma_\alpha  \gamma_m \Delta_\alpha ^4\right),\\
D&=g^2 \Delta_\alpha  \cosh (2 \theta ) \left(\gamma_\alpha  \gamma_m \Delta_\alpha ^3-\gamma_\alpha  \Delta_m^3 (2 \gamma_\alpha +\gamma_m)+\Delta_\alpha  \Delta_m^2 (2 \gamma_\alpha +\gamma_m)^2-\gamma_\alpha  \Delta_\alpha ^2 \Delta_m (\gamma_\alpha +2 \gamma_m)\right .\\
&\left .+\gamma_m \Delta_\alpha  (\gamma_\alpha +\gamma_m) \left(\gamma_\alpha ^2+\gamma_m^2+3 \gamma_\alpha  \gamma_m+g^2\right)-\gamma_\alpha  \Delta_m (\gamma_\alpha +\gamma_m) \left(\gamma_\alpha ^2+\gamma_m^2+\gamma_\alpha  \gamma_m-g^2\right)\right),\\
A'&=\gamma_m \left(\gamma_\alpha ^2+\Delta_\alpha ^2\right) (\gamma_\alpha +\gamma_m) \left(\gamma_m^2+\Delta_m^2\right) \left((\gamma_\alpha +\gamma_m)^2+(\Delta_m-\Delta_\alpha )^2\right),\\
B'&=g^4 (\gamma_\alpha +\gamma_m) \left(\gamma_\alpha  \Delta_m^2+\gamma_m^3+2 \gamma_\alpha  \gamma_m^2+\gamma_m \left(\gamma_\alpha ^2+\Delta_\alpha ^2+\Delta_m^2-\Delta_\alpha  \Delta_m\right)\right),\\
C'&=g^2 \left(\gamma_m \Delta_\alpha  \Delta_m \left(\gamma_\alpha ^3+\gamma_m^3+\gamma_m \Delta_m^2\right)+\gamma_\alpha  \left(\Delta_m^2 (\gamma_\alpha +\gamma_m) \left(\gamma_\alpha ^2+3 \gamma_m^2+3 \gamma_\alpha  \gamma_m\right)+2 \gamma_m^2 (\gamma_\alpha +\gamma_m)^3+\gamma_m \Delta_m^4\right) \right., \\
&\left . +\gamma_\alpha  \Delta_\alpha ^2 \left(\gamma_\alpha  \Delta_m^2+2 \gamma_m^2 (\gamma_\alpha +\gamma_m)\right)+\gamma_\alpha  \gamma_m \Delta_\alpha ^3 \Delta_m\right),\\
D'&=g^2 \Delta_m \cosh (2 \theta ) \left(-\gamma_m \Delta_\alpha ^3 (\gamma_\alpha +2 \gamma_m)+\gamma_\alpha  \gamma_m \Delta_m^3+\Delta_\alpha ^2 \Delta_m (\gamma_\alpha +2 \gamma_m)^2 \right .,\\
&\left .-\gamma_m \Delta_\alpha  \left(\Delta_m^2 (2 \gamma_\alpha +\gamma_m)+(\gamma_\alpha +\gamma_m) \left(\gamma_\alpha ^2+\gamma_m^2+\gamma_\alpha  \gamma_m-g^2\right)\right)+\gamma_\alpha  \Delta_m (\gamma_\alpha +\gamma_m) \left(\gamma_\alpha ^2+\gamma_m^2+3 \gamma_\alpha  \gamma_m+g^2\right)\right),\\
Z_1&=g^6 (\gamma_\alpha +\gamma_m)^2 \left((\gamma_\alpha +\gamma_m)^2+\Delta_\alpha ^2+\Delta_m^2\right)\\
&+\gamma_\alpha  \gamma_m \left(\gamma_\alpha ^2+\Delta_\alpha ^2\right) \left(\gamma_m^2+\Delta_m^2\right) \left(\left((\gamma_\alpha +\gamma_m)^2+\Delta_m^2\right)^2+2 \Delta_\alpha ^2 (\gamma_\alpha +\gamma_m-\Delta_m) (\gamma_\alpha +\gamma_m+\Delta_m)+\Delta_\alpha ^4\right),\\
Z_2&=g^4 \left(2 \Delta_\alpha ^2 \left(2 \gamma_\alpha  \gamma_m (\gamma_\alpha +\gamma_m)^2-\Delta_m^2 \left(\gamma_\alpha ^2+\gamma_m^2+3 \gamma_\alpha  \gamma_m\right)\right)+\gamma_\alpha  \gamma_m \Delta_\alpha ^4 \right .\\
&\left .+\gamma_\alpha  \gamma_m \left(4 \Delta_m^2 (\gamma_\alpha +\gamma_m)^2+3 (\gamma_\alpha +\gamma_m)^4+\Delta_m^4\right)\right),\\
Z_3&=g^2 \left(\Delta_\alpha ^4 \left(\gamma_m^2 \left(3 \gamma_\alpha ^2+\gamma_m^2+2 \gamma_\alpha  \gamma_m\right)+\Delta_m^2 (\gamma_\alpha +\gamma_m)^2\right) \right.\\
&\left .+\Delta_\alpha ^2 \left(2 \Delta_m^2 \left(\gamma_\alpha ^2+\gamma_m^2\right) \left(\gamma_\alpha ^2+\gamma_m^2+3 \gamma_\alpha  \gamma_m\right)+\gamma_m^2 (\gamma_\alpha +\gamma_m)^2 \left(6 \gamma_\alpha ^2+\gamma_m^2+2 \gamma_\alpha  \gamma_m\right)+\Delta_m^4 (\gamma_\alpha +\gamma_m)^2\right) \right .\\
&\left. +\gamma_\alpha ^2 \left(\Delta_m^4 \left(\gamma_\alpha ^2+3 \gamma_m^2+2 \gamma_\alpha  \gamma_m\right)+\Delta_m^2 \left(\gamma_\alpha ^2+6 \gamma_m^2+2 \gamma_\alpha  \gamma_m\right) (\gamma_\alpha +\gamma_m)^2+3 \gamma_m^2 (\gamma_\alpha +\gamma_m)^4\right)\right),\\
Z_4&=2 g^2 \Delta_\alpha  \Delta_m \cosh (2 \theta ) \left (\Delta_\alpha ^2 \left(\Delta_m^2 \left(\gamma_\alpha ^2+\gamma_m^2+4 \gamma_\alpha  \gamma_m\right)+\gamma_m (\gamma_m-2 \gamma_\alpha ) (\gamma_\alpha +\gamma_m)^2\right) \right .,\\
&\left .+\gamma_\alpha  \left(-\gamma_m (\gamma_\alpha +\gamma_m)^2 \left(\gamma_\alpha ^2+\gamma_m^2+\gamma_\alpha  \gamma_m\right)+\Delta_m^2 (\gamma_\alpha -2 \gamma_m) (\gamma_\alpha +\gamma_m)^2-\gamma_m \Delta_m^4\right)\right .,\\
&\left .-\gamma_\alpha  \gamma_m \Delta_\alpha ^4+g^4 (\gamma_\alpha +\gamma_m)^2-g^2 (\gamma_\alpha +\gamma_m)^2 \left(\gamma_\alpha ^2+\gamma_m^2+\Delta_\alpha ^2+\Delta_m^2\right) \right), \\
&-2\left (g^2 \Delta_\alpha  \Delta_m \right )^2(\gamma_\alpha +\gamma_m)^2 \cosh (4 \theta ).
\end{aligned}
\end{equation}
\end{widetext}

To understand the analytical results, we first consider the case that photons are dissipationless, i.e. $\gamma_\alpha=0$, then the magnon density can be simplified into the form,
\begin{widetext}
\begin{equation}
\langle m^\dagger m \rangle = \frac{2g^2\Delta_\alpha^2(g^2+\gamma_m^2+\Delta_m^2)\mathrm{cosh}^2\theta \mathrm{sinh}^2\theta}{(\Delta_\alpha^2+\gamma_m^2+\Delta_m^2+2\Delta_\alpha \Delta_m \mathrm{cosh}2\theta )(g^4+(\gamma_m^2+\Delta_m^2)\Delta_\alpha^2-2g^2\Delta_m \Delta_\alpha \mathrm{cosh} 2\theta)}.
\end{equation}
\end{widetext}

Since we are interested in the off-resonant excitation of magnons, i.e. $\Delta_\alpha \gg g \gg \Delta_m,\gamma_m$, the magnon density is further reduced to,
\begin{widetext}
\begin{equation}
\langle m^\dagger m \rangle \approx \frac{2g_c^2g_s^2}{g^4+\gamma_m^2\Delta_\alpha^2+\Delta_\alpha^2(\Delta_m^2-2g^2\Delta_m \mathrm{cosh}(2\theta)/\Delta_\alpha)},
\label{approx}
\end{equation}
\end{widetext}
where $g_c=g \sinh \theta, g_s =g\cosh \theta$. This is the formula used in the main text.

\section{Unified understanding of the off-resonant excitation}
\begin{figure}
\centering
\includegraphics[width=0.45\textwidth]{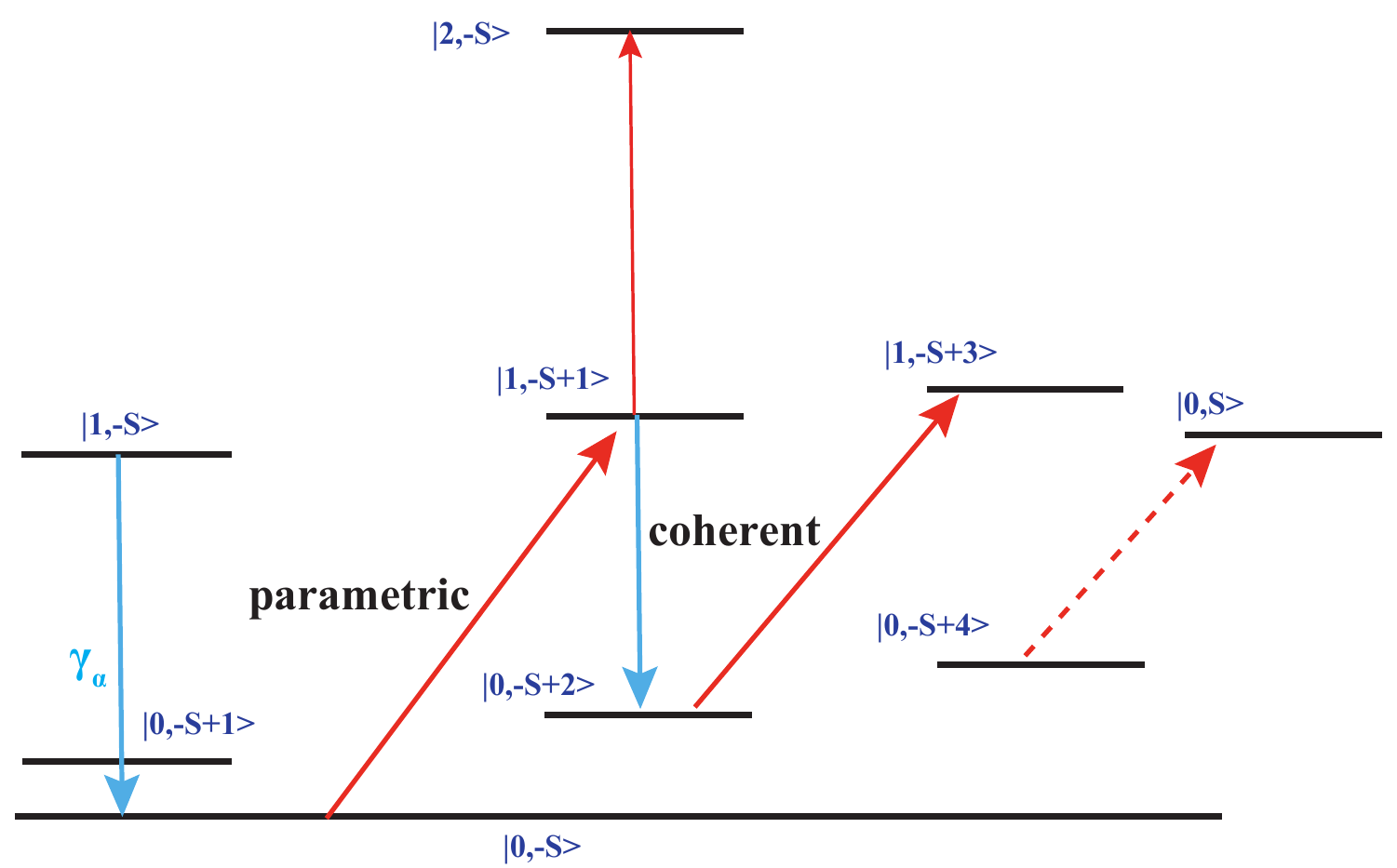}
\caption{Schematic of the energy level diagram when the Mode 2 is a low frequency spin mode.}
\label{spin_level}
\end{figure}

In this section, we summarize the off-resonant behavior of a wide class of quantum optical systems in Table \ref{tab2}.
Qualitatively, our physical picture presented in Fig. 2(e) of the main text provide a unified understanding of all these observations.
For Heisenberg model and $s$-$d$ model, the low frequency spin mode has a bounded Hilbert space, as shown in Fig. \ref{spin_level}. The magnons will first be
largely excited to occupy the Fock state $|2\rangle, |4\rangle,...,|S\rangle$ and then the system will keep oscillating around these states. Once dissipation is introduced, the system may stabilize at a new
 steady state. If the low frequency mode is a boson, its Hilbert space is unbounded and thus the number of excited particles
will keep increasing and diverge. A sufficiently strong dissipation can stabilize the system and sustain a finite excitation, as discussed in the main text.
On the other hand, the parametric excitations in these models can be neglected on-resonance based on rotational wave approximation and the underlying physics will be changed.
A comprehensive study of the off-resonant behaviors in these models including the critical frequency and influence of various types of spin dephasing will be published elsewhere.

\begin{table*}
\centering
\caption{Summary of the on-resonant and off-resonant excitation in the absence of dissipation. $\mathcal{H}=\omega_1 \Omega(\mathcal{O}_1) + \omega_2 \Omega(\mathcal{O}_2)+g(\mathcal{O}_1^\dagger + \mathcal{O}_1)(\mathcal{O}_2^\dagger + \mathcal{O}_2)$, $\Omega(\mathcal{O}_i)=\mathcal{O}_{i,z}$ for spin operator and $\mathcal{O}_i^\dagger \mathcal{O}_i$ for bosonic operators. Parameters are $\omega_1=1.0,g=0.1,S=16$. ST for steady state, OS for oscillation.}
\begin{tabular}{c|c|c|c|l|c|l}
 \hline
 \hline
  Mode 1 & Mode 2 & Model name & Off-resonance&Behavior of Mode 2&On-resonance& Behavior of Mode 2\\
\hline
 $a_1$ & $a_2$  & Boson-Boson & $\omega_2=0.002 \omega_1$ & Diverge &$\omega_2=\omega_1$ & OS $\langle a_2^\dagger a_2 \rangle \sim0.005$\\
 $a$   &$\sigma$ & Rabi model& $\omega_2=0.002 \omega_1$ & OS $\langle \sigma_{z} \rangle =-0.48$ & $\omega_2=\omega_1$ & OS $\langle \sigma_z \rangle =-0.497$ \\
 \hline
 $\sigma$ & $a$  & Rabi model& $\omega_2=0.002 \omega_1$  & Diverge & $\omega_2=\omega_1$  & OS $\langle a^\dagger a \rangle =0.006$\\
 $\sigma_1$ &$\sigma_2$ & Heisenberg X & $\omega_2=0.002\omega_1$  & OS $\langle \sigma_{2z} \rangle =-0.48$ & $\omega_2=\omega_1$ & OS $\langle \sigma_{2z} \rangle =-0.495$\\
 $\sigma$   &$S$ & $s$-$d$ model* $\cite{notesd}$&$\omega_2=0.002\omega_1$   & OS $\langle S_z \rangle =-2.03$&$\omega_2=\omega_1$  & OS $\langle S_{z} \rangle =-15.71$\\
 \hline
 $S$   &$\sigma$ & $s$-$d$ model*& $\omega_2=0.002\omega_1$  & OS $\langle \sigma_z \rangle =-0.21$ &$\omega_2=\omega_1$  & OS $\langle \sigma_z \rangle =-0.38 $\\
 $S_1$   &$S_2$ & Heisenberg X & $\omega_2<g^2/\omega_1$  & Tilted OS & $\omega_2=\omega_1$ & No tilt OS\\
 \hline
 \hline
 $x_1$&$x_2$ & coupled oscillators & $\omega_2<g^2/\omega_1$ &Diverge & $\omega_2=\omega_1$& OS\\
 \hline
 \hline
\end{tabular}
\label{tab2}
\end{table*}

\section{Calculation of the quantum steering and Wigner distribution}
To quantify the steerability existing between magnon mode and photon mode, we first recall the covariance matrix of the hybrid system defined as,
\begin{equation}
\mathbf{V}=\left ( \begin{array}{cc}
            \mathbf{V}_m & \mathbf{C} \\
            \mathbf{C}^T & \mathbf{V}_\alpha
          \end{array}
\right ),
\end{equation}
where $V_{ij}=\langle R_i R_j +R_j R_i\rangle/2, \mathbf{R}=(x_m,p_m,x_\alpha,p_\alpha)$. All these elements can be analytically calculated based on the solution of Eq. (\ref{leq10}) as,
\begin{subequations}
\begin{align}
&\langle x_\nu x_\nu \rangle = n_\nu + Re(\langle \nu\nu \rangle)+\frac{1}{2},\\
&\langle p_\nu p_\nu \rangle = n_\nu - Re(\langle \nu\nu \rangle)+\frac{1}{2},\\
&\langle x_\alpha x_m \rangle = \langle x_m x_\alpha \rangle=Re(\langle \alpha m  \rangle) + Re(\langle \alpha^\dagger m  \rangle),\\
&\langle p_\alpha p_m \rangle = \langle p_m p_\alpha \rangle =-Re(\langle \alpha m  \rangle) + Re(\langle \alpha^\dagger m  \rangle),\\
&\langle x_\alpha p_m \rangle = \langle p_m x_\alpha \rangle =Im(\langle \alpha m  \rangle) + Im(\langle \alpha^\dagger m  \rangle),\\
&\langle x_m p_\alpha \rangle = \langle p_\alpha x_m \rangle =Im(\langle \alpha m  \rangle) - Im(\langle \alpha^\dagger m  \rangle),
\end{align}
\end{subequations}
where $\nu = m, \alpha$.

For a two-mode Gaussian state as studied here, it has been shown that \cite{Reid1989,Kogias2015} the sufficient and necessary condition, for the two-mode steering from magnon to photon can be verified using the inequality,
\begin{equation}
E_{m|\alpha}=V_{x_\alpha|x_m}V_{p_\alpha|p_m}=4\det V/\det V_m<1,
\end{equation}
where $V_{x_\alpha|x_m}$($V_{p_\alpha|p_m}$) is minimum inferred variance of the photon quadrature $x_\alpha (p_\alpha)$ provided that magnon is measured at $x_m$ ($p_m$).

To obtain the Wigner distribution of the steady magnon state shown in Fig. 3(b) and (c) of the main text, we recall the Lindblad master equations of the hybrid system,
\begin{equation}
\frac{d \rho}{dt}=-i [\mathcal{H},\rho]  + \mathcal{L}^{(m)} \rho+ \mathcal{L}^{(\alpha)} \rho,
\label{numericeq}
\end{equation}
where $\mathcal{L}^{(m)}\rho= \sum_{n=1,2} [C_n\rho C_n^\dagger-(\rho C_n^\dagger C_n+ C_n^\dagger C_n\rho)/2]$, with
 $C_1=\sqrt{(n_\mathrm{th}+1)\gamma_m} m$ and $C_2=\sqrt{n_\mathrm{th}\gamma_m} m^\dagger$ that describe the process of magnon annihilation and creation, respectively, $n_\mathrm{th}$ is the magnon population in thermal equilibrium. Similar definition follows for the photonic part $\mathcal{L}^{(\alpha)} \rho$. The density matrix of the steady state is obtained by numerically solving Eq. (\ref{numericeq}) by a home-made code and further verified using the open-source package \textit{QuTip} \cite{qutip}. The Wigner function is calculated in \textit{QuTip} following its definition,
 \begin{equation}
 W(x,p)=\frac{1}{\pi} \int \langle x+\zeta| \rho |x-\zeta \rangle e^{-2ip\zeta} d \zeta.
 \end{equation}

\section{Experimental parameters of hybrid magnet-light systems}
In this section, we list the experimental values of the magnon-photon coupling and the resonant frequencies in Table \ref{tab1}. Note that the $g/\omega_m$ ranges from $0.002$ to $0.23$, while the value 0.03 used in the main text falls into this regime.
Most experiments used yttrium iron garnet with a damping constant $10^{-5} \sim 10^{-3}$, while the value used in the theoretical prediction is $10^{-4}$.

\begin{table}
\caption{List of the magnetic parameters in the typical experimental setups. $\omega_m$ is the frequency of cavity mode,
while $g$ is the coupling strength between the cavity mode and magnon mode.}
\begin{tabular}{l|c|c|c|c}
        \hline
        \hline
            Experiments & year & $\omega_m$ (GHz) &$g$ (MHz)&$g/\omega_m$\\
             \hline
            CPW Resonator $\cite{Huebl2013}$  & 2013 & $\sim 6$ & 450&0.075\\
            Reentrant cavity  $\cite{Gor2014}$ &2014&  12.9& 124 &0.01\\
            Rectangular cavity(Al) $\cite{Bai2015}$ & 2015 & 11.5 & 180 &0.016\\
            \hline
            Rectangular cavity (Cu) $\cite{Wang2018}$& 2018 & 10.1 & 41 &0.004\\
            Fabry-Perot cavity $\cite{Harder2018}$ & 2018& 13.2& 39 &0.003\\
            \hline
            ISSR resonator  $\cite{Bohi2019}$ & 2019 & $\sim 4$ & 180&0.045\\
            Cross-line cavity $\cite{YPWang2019}$ & 2019& 4.7 & 7.9 &0.002\\
            photonic crystal $\cite{Zhang2019apl}$ & 2019&$\sim 9.0$ & 2100 &0.23\\
             \hline
           \end{tabular}
\label{tab1}
\end{table}

\section{Classical counterparts}

In this section, we show the absence of the off-resonant excitation in typical classical systems, including two coupled classical harmonic oscillators (CHO-CHO) and two coupled spins (spin-spin model), even though they are described by Hamiltonian resembling Eq. (3) of the main text.
\subsection{CHO-CHO model}
The first model is the two harmonic oscillators coupled via a spring, which is described by the classical Hamiltonian,
\begin{equation}
\mathcal{H}_{cl}=\frac{1}{2}\sum_{i=1}^2(m_i\dot{x}_i^2 + m_i\omega_i^2 x_i^2)+k(x_1-x_2)^2,
\end{equation}
where $m_i$ is the mass, $\omega_i$ is the intrinsic frequency of the two oscillators, and $k$ is the stiff of the spring connected the two oscillators. By introducing the quadratures $x_i=(a_i+a_i^\dagger)/\sqrt{2m_i\tilde{\omega}_i},p_i=-i(a_i-a_i^\dagger)\sqrt{m_i\tilde{\omega}_i/2}$, it is straightforward to obtain the quantum mechanical counterpart of this Hamiltonian as,
\begin{equation}
\mathcal{H}_{qm}=\sum_{i=1}^2 \tilde{\omega}_i a_i^\dagger a_i + g(a_1^\dagger a_2 + a_1^\dagger a_2^\dagger + h.c.),
\end{equation}
where $\tilde{\omega}_i^2 = \omega_i^2 + 2k/m_i,g=-k/\sqrt{m_1m_2 \tilde{\omega}_1 \tilde{\omega}_2}$. This resembles Eq. (3) of main text, but with $g_c=g_s=g$ and revision of the eigen-frequency by the coupling constant. It is straightforward to obtain the eigen-spectrum as,
\begin{equation}
\tilde{\omega}^2=\frac{1}{2}\left(\tilde{\omega}_1+\tilde{\omega}_2 \pm \sqrt{(\tilde{\omega}_1+\tilde{\omega}_2)^2 +16g^2 - 4 \tilde{\omega}_1 \tilde{\omega}_2}\right ).
\label{qmg}
\end{equation}
Due to the significant modification of natural frequency $\tilde{\omega}_i$ by the mutual coupling $g$, $\tilde{\omega}^2$ is always larger than zero, and no anomaly appears. Similar arguments apply to the two inductively coupled LC circuits.

\subsection{Spin-spin model}
The second model is the two ferromagnetically coupled macrospins $\mathbf{S}_1$ and $\mathbf{S}_2$, governed by the Hamiltonian,
\begin{equation}
\mathcal{H}_{cl}=-S_{1,z} B_1 - S_{2,z} B_2 - J S_{1x} S_{2x}.
\label{2spin}
\end{equation}
The ground state can be obtained by rewriting the Hamiltonian in spherical coordinate $\mathcal{H}_{cl}=-B_1\cos \theta_1 - B_2 \cos \theta_2 - J  \sin \theta_1 \sin \theta_2$ and minimizing the total energy with respect to the tilting angle of spin as,
\begin{widetext}
\begin{subequations}
\begin{align}
&(1) \mathrm{When~~} B_1B_2>J^2, \theta_1=\theta_2=0\label{noflop},\\
&(2)\mathrm{When~} B_1B_2<J^2, \tan \theta_1=\sqrt{\frac{J^4-(B_1B_2)^2}{B_1^2(B_2^2+J^2)}}, J\sin \theta_2=B_1 \tan \theta_1. \label{flop}
\end{align}
\end{subequations}
\end{widetext}
This implies that as we tune the magnitude of $B_2$ below a threshold $J^2/B_1$, the system will flop to a non-collinear group state.
When $B_1B_2>J^2$, using standard Bogliubov transformation around Eq. (\ref{noflop}), i.e. $\mathbf{S}_1 =\mathbf{S}_2 =e_z$, Eq. (\ref{2spin}) can be diagonalized into the same form as Eq. (\ref{qmg}), with $\tilde{\omega}_i=B_i, g=J/2$,  obviously, the eigenvalues are always reals. When $B_1B_2>J^2$ or $g>\tilde{\omega}_1 \tilde{\omega}_2/4$, the eigenvalues are imaginary, but the ground states $\mathbf{S}_1 =\mathbf{S}_2 =Se_z$ become unstable in this parameter regime. To find the true spectrum, we have to perform expansion near the ground state Eq. (\ref{flop}). For this purpose, we define a new coordinate $x'y'z'$, where $z'$ align along the direction of the spin and $y'=y$, then we have
\begin{equation}
\left ( \begin{array}{c}
          S_{i,x} \\
          S_{i,y} \\
          S_{i,z}
        \end{array}
\right )=\left (
\begin{array}{ccc}
  \cos \theta_i & 0 & \sin \theta_i\\
  0 & 1 & 0 \\
  -\sin \theta_i & 0 & \cos \theta_i
\end{array}\right )
\left ( \begin{array}{c}
          S_{i,x'} \\
          S_{i,y'} \\
          S_{i,z'}
        \end{array}
\right ),
\end{equation}
where $S_{i,z'}=S-a_i^\dagger a_i,S_{i,x'}=\sqrt{2S}/2(a_i^\dagger +a_i),S_{i,y'}=-i\sqrt{2S}/2(a_i^\dagger -a_i)$. By substituting these transformations into the original Hamiltonian Eq. (\ref{2spin}), we obtain
\begin{widetext}
\begin{equation}
\mathcal{H}_{QM}= (B_1 + JS^2 \sin \theta_1 \sin \theta_2)a_1^\dagger a_1 + (B_2 + JS^2 \sin \theta_1 \sin \theta_2)a_2^\dagger a_2-\frac{JS}{2} \cos \theta_1 \cos \theta_2 (a_1^\dagger + a_1)(a_2^\dagger + a_2).
\end{equation}
\end{widetext}
Now the critical condition to have the complex spectrum becomes,

\begin{widetext}
\begin{equation}
(B_1 + JS^2 \sin \theta_1 \sin \theta_2)(B_2 + JS^2 \sin \theta_1 \sin \theta_2)<(JS\cos \theta_1 \cos \theta_2)^2.
\label{cond}
\end{equation}
\end{widetext}
Considering the premise condition Eq. (\ref{flop}) for the ground state, it is straightforward to numerically verify that this inequality cannot be satisfied. One can also reach this conclusion analytically, by noting that $B_2 \ll J \ll B_1$ in this spin-flop regime, which implies $\tan \theta_1 \approx \sin \theta_1 \approx J/B_1$, $\theta_2\approx \pi/2$. Therefore the right-hand side of (\ref{cond}) is zero while the left-hand side is larger than zero, then the inequality cannot be satisfied. Hence the spectrum of the system at the spin-flopped phase is always real.

\begin{figure}[H]
	\centering
	\includegraphics[width=0.48\textwidth]{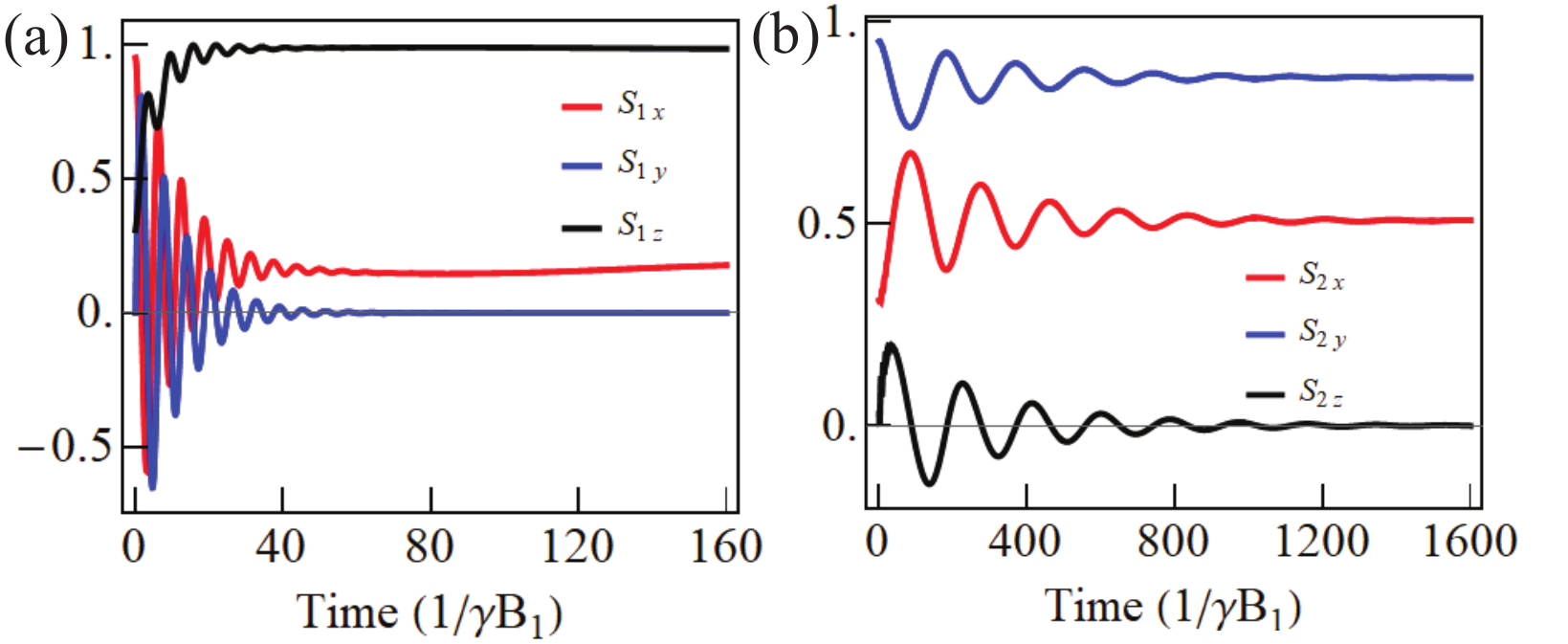}
	\caption{Time evolution of the spin orientation for the first spin (a) and second spin (b), respectively. Parameters are $B_1=1.0, B_2=0.02,J=0.2,\alpha=0.1$.}
	\label{spin_time}
\end{figure}

We also verify the absence of the off-resonant excitation in the spin-spin model by numerically solving the LLG equation,
\begin{equation}
\frac{\partial \mathbf{S}_i}{\partial t}=-\mathbf{S}_i\times \mathbf{H}_{\mathrm{eff},i} + \alpha \mathbf{S}_i \times \frac{\partial \mathbf{S}_i}{\partial t},
\label{llg}
\end{equation}
where $\mathbf{H}_{\mathrm{eff},i}$ is the effective field acting on spin $\mathbf{S}_i$, defined as $\mathbf{H}_{\mathrm{eff},i}=-\delta \mathcal{H}_{cl}/\delta \mathbf{S}_i$, $\alpha$ is the damping constant which represents how fast the spin moves toward the effective field. The evolution of the spin orientation when the frequency of the second spin is reduced below the limit $J^2/B_1$ is shown in Fig. \ref{spin_time}. The steady state of the coupled systems deviate from the $+z$ state and stabilize at a canted state, while there are no finite oscillations of spins in the steady state. This is fully consistent with the results obtained by analyzing the spectrum.

\subsection{Artificial CHO-CHO}
\begin{figure}[H]
	\centering
	\includegraphics[width=0.48\textwidth]{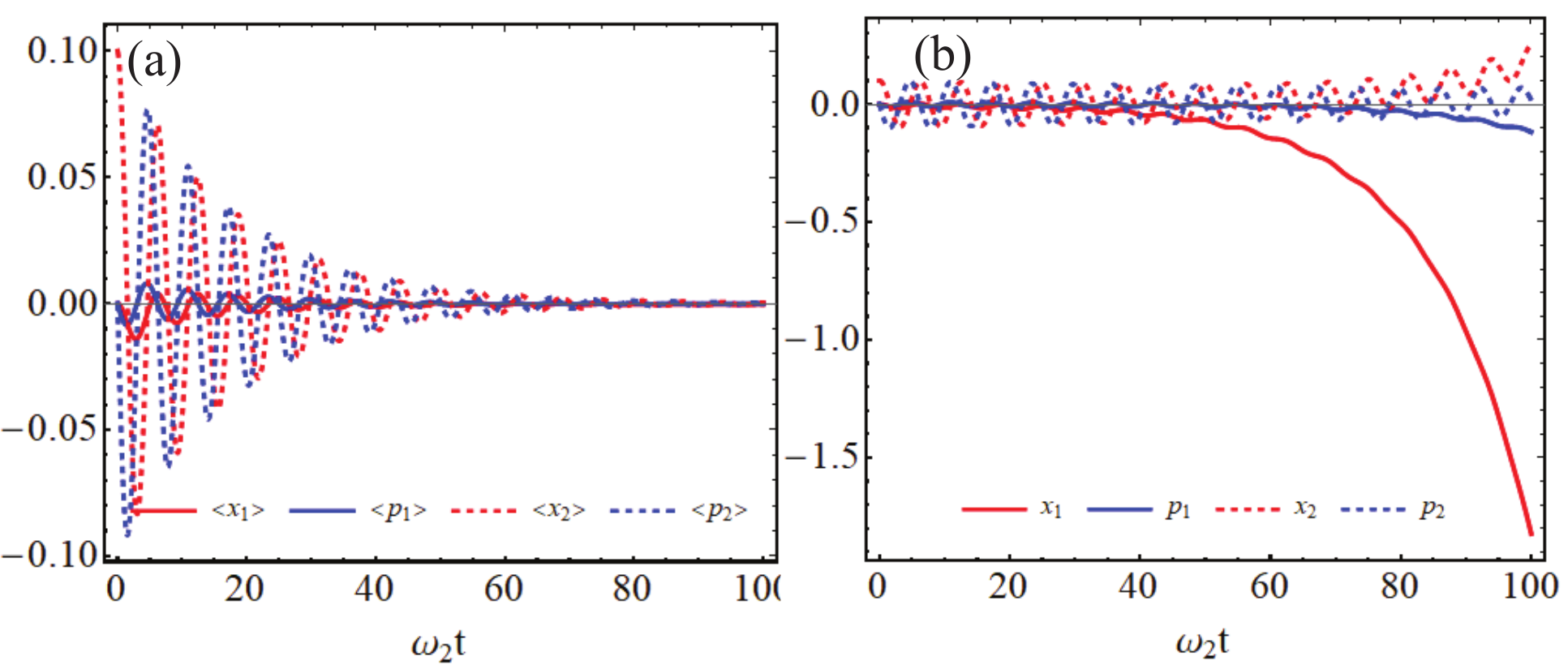}
	\caption{(a) Time evolution of the order parameters in quantum mechanical system (a) and classical system (b), respectively. Parameters are $\omega_1=0.003\omega_2,g=0.1\omega_2$, $\gamma_1=\gamma_2=0.1\omega_2$  for (a) and $\mu_1=\mu_2=0.1\omega_2$ for (b).}
	\label{sfig1}
\end{figure}

In Sec. IVA, we show that the coupling between two harmonic oscillators will modify the self-frequency of the systems and thus avoid the system to go into the anomalous regime. Here we further ask what if the coupling does not modify the self-eigenfrequency? One artificial coupling that can realize this goal is $gx_1\cdot x_2$. The dynamic equation in terms of Newton law can be rewritten as  $d\mathbf{R}/dt=\mathbf{M}_{\mathrm{cl}} \cdot \mathbf{R}$, where $\mathbf{R}=(x_1,p_1,x_2,p_2)^T$ and the dynamic equation $\mathbf{M}_{\mathrm{cl}}$ reads,
\begin{equation}
\mathbf{M}_{\mathrm{cl}}=\left ( \begin{array}{cccc}
            0 & 1 &0&0 \\
            -\omega_1^2 & -\mu_1 &-g&0 \\
            0 & 0 &0&1 \\
            -g & 0 &-\omega_2^2&-\mu_2 \\
          \end{array}
\right ),
\end{equation}
To make a fair comparison with the quantum mechanical case, we define the canonical transformation $x_i=1/\sqrt{2\omega_1}(a_i+a_i^\dagger),p_i=\sqrt{\omega_1/2i}(a_i-a_i^\dagger)$ to transform the quantum dynamics matrix in Eq. (4) of the main text to a similar form to its classical counterpart, except the extra dissipation on the coordinate dimension, i.e. $d\mathbf{R}/dt=\mathbf{M}_{\mathrm{qm}} \cdot \mathbf{R}$ with $\mathbf{R}=(\langle x_1 \rangle,\langle p_1 \rangle,\langle x_2 \rangle,\langle p_2\rangle)^T$, and dynamic matrix,
\begin{equation}
\mathbf{M}_{\mathrm{qm}}=\left ( \begin{array}{cccc}
            -\gamma_1 & 1 &0&0 \\
            -\omega_1^2 & -\gamma_1 &-g&0 \\
            0 & 0 &-\gamma_2&1 \\
            -g & 0 &-\omega_2^2&-\gamma_2 \\
          \end{array}
\right ).
\end{equation}
By numerically solving the dynamic equation in both the classical and quantum mechanical case, we obtain the results shown in Fig. \ref{sfig1}.

Clearly, the system described by quantum mechanical equation can gradually evolve to a steady state (Fig. \ref{sfig1}(a)) while it diverges slowly in the classical case (Fig. \ref{sfig1}(b)), even if they share exactly the same parameters. This is due to the absence of dissipation in the $x$ coordinate of the oscillator in the classical case. Then the oscillator 1 can slowly drift away from its equilibrium position and finally diverge (red line in Fig. \ref{sfig1}(b)).

Through these examples, we find that the coupling of two classical modes can delicately modify the mode frequencies to guarantee that the hybrid eigen-spectrum is real. When the intrinsic mode dissipation is introduced, the hybrid modes gradually evolve to the classical ground state, without any particle fluctuation. In quantum mechanics, the ground state becomes an entangled squeezed state without classical correspondence and the (quasi-)particles acquire a finite occupation in this squeezed vacuum.


\begin{thebibliography}{}
\bibitem{notepic} This intuitive picture provides a unified understanding of the off-resonant behavior of a wide class of physical systems, such as Rabi model, $s$-$d$ model, classical Heisenberg model etc. Depending on whether the low frequency party has a bounded Hilbert space (such as a spin state) or unbounded Hilbert space (such as bosonic mode), the system may stabilize at a new ground state or sustain a finite excitation.
\bibitem{note1} Note here the squeezing of the magnon mode is not strictly along the $x$ and $p$ direction
due to its sensativity on the detuning between magnon and photon modes. In general, there should exist an optimized angle $x_\theta = 1/\sqrt{2}(m e^{-i\theta} + m^\dagger e^{i\theta}),p_\theta = 1/\sqrt{2}i(m e^{-i\theta} - m^\dagger e^{i\theta})$ that maximizes the squeezing.
\bibitem{notesd} The original $s$-$d$ exchagne coupling descries the ineraction between conduction electrons ($\sigma$) and the localized magnetic
moments (spin $\mathbf{S}$) as $g\sigma \cdot \mathbf{S}$. Here we mark a star in this model becasue we consider a reduced type of interaction
$g(\sigma^+ + \sigma^-) \cdot(S^+ + S^-)$.
\end{thebibliography}
\end{document}